\documentclass[reprint,prb, amsmath,amssymb,aps]{revtex4-1}
\usepackage[dvipdfmx]{graphicx}
\usepackage{amsthm,bm}
\usepackage{comment}

\begin{document}

\preprint{APS/123-QED}

\title{Effects of Coulomb interaction \\ on photon-assisted current noises through a quantum dot}

\author{Takafumi J. Suzuki}\author{Takeo Kato}
\affiliation{Institute for Solid State Physics, University of Tokyo, Kashiwa, Chiba, 277-8581, Japan}

\date{\today}


\begin{abstract}        

We study photon-assisted transport in a single-level quantum dot system under a
periodically oscillating field.
Photon-assisted current noises in the presence of the Coulomb interaction are calculated based on a gauge-invariant formulation of time-dependent transport.
We derive the vertex corrections within the self-consistent Hartree-Fock approximation in terms the Floquet-Green's functions (Floquet-GFs), and examine the effects of the Coulomb interaction on the photon-assisted current noises.
Moreover, we introduce a concept of an effective temperature to characterize nonequilibrium properties under the influence of the AC field.
The vertex corrections are suppressed by the rise of the effective temperature, whereas characteristic resonant structures appear in the frequency spectra of the vertex corrections.
The present result provides a useful viewpoint for understanding photon-assisted transport in interacting electron systems.

\end{abstract}


\maketitle


\section{Introduction}

Photon-assisted transport through mesoscopic conductors has attracted much attention because the external fields open up additional transport channels via photon absorption and emission.~\cite{PhysRev.129.647,platero2004photon,kohler2005driven}
Nonperturbative effects of the time-dependent fields significantly modify the quantum nature of transport processes, and these effects have been studied from various viewpoints, such as the coherent destruction of tunneling,~\cite{PhysRevLett.67.516} the nonstationary Aharonov-Bohm effect,~\cite{lesovik1994noise} and so on.
Photon-assisted transport has also been utilized in various applications for electronic devices, e.g., classical current sources,~\cite{Pekola13} operation of solid state quantum bits,~\cite{van2002electron} and on-demand generation of quantum excitations.~\cite{feve2007demand,dubois2013minimal}

In recent years, much studies have revealed that current noises provide significant information about the microscopic processes involved in photon-assisted transport.~\cite{ThierryReview,blanter2000shot}
Levitov and Lesovik pointed out that photon-assisted current noises can be used to detect the phase of the transmission amplitudes induced by the external AC field.~\cite{lesovik1994noise}
In subsequent theoretical work, the coherent and spectroscopic nature of the photon-assisted current noise of noninteracting electrons has been studied based on the scattering approach~\cite{hammer2011quantum} and the Green's function (GF) approach.~\cite{PhysRevLett.90.210602}
A detection scheme for finite-frequency current noises under an AC field using a resonant circuit has been proposed theoretically.~\cite{chevallier2010detection}
Photon-assisted current noises have also been measured in various systems such as diffusive metals,~\cite{schoelkopf1998observation} diffusive normal metal-superconductor junctions,~\cite{kozhevnikov2000observation} quantum point contacts,~\cite{reydellet2003quantum} and tunnel junctions.~\cite{gabelli2008dynamics,gasse2013observation} 
Recently, the time-resolved current noises have been measured to evaluate the quantum purity of electrons emitted from on-demand electron sources.~\cite{Olkhovskaya08,bocquillon2013coherence,dubois2013minimal}

Although the scattering theory~\cite{Pedersen98,blanter2000shot} has clarified the properties of photon-assisted current noises of noninteracting electrons, it is of limited use to describe the effects of the Coulomb interaction.
A quantum dot (QD) is a typical system in which the Coulomb interaction significantly affects the transport properties. 
This raises the question of whether qualitative features of the photon-assisted current noises obtained in noninteracting electron systems should or should not be altered in the presence of the Coulomb interaction.
One sophisticated way to tackle this problem is a perturbative expansion with respect to the Coulomb interaction.
For a reliable description of nonequilibrium transport in the presence of the Coulomb interaction, we need to carefully consider the charge conservation law, which is equivalent to gauge invariance.
Hershfield showed that the vertex corrections are essential for satisfying the gauge invariance of current noises, and derived their explicit expressions for zero-frequency noises under stationary bias voltages within the self-consistent Hartree-Fock (SCHF) approximation.~\cite{PhysRevB.46.7061}
Recently, Ding and Dong discussed the charge-conserving approximation~\cite{PhysRev.124.287,PhysRev.127.1391} for time-dependent transport quantities, and studied the finite-frequency current noises of the same system.~\cite{ding2013finite}
We note that a similar calculation using the charge-conserving approximation was performed for a QD system with superconducting leads.~\cite{rech2012current}
However, the vertex corrections of the current noises through a QD system under time-dependent external fields have not yet been studied.

The purpose of this paper is to study effects of the Coulomb interaction on the photon-assisted current noises using the gauge-invariant approximation scheme.
We consider a single-level QD system under a periodically-oscillating external field, and derive the explicit expressions of the vertex corrections of the photon-assisted current noises within the SCHF approximation. 
Using these expressions, we examine the features of the vertex corrections under the AC field.

This paper is organized as follows:
In Sec.~\ref{chap:formulation}, we provide our model and describe a gauge-invariant formulation of the time-dependent transport phenomena.
We present the expressions of the vertex functions within the SCHF approximation, and introduce Floquet-GFs to describe transport in the periodically driven system.
In Sec.~\ref{chap:PAT}, we introduce the generalized distribution, which captures characteristic features of the AC field.
In Sec.~\ref{chap:noise_non}, an effective temperature is introduced to understand the properties of zero-frequency current noises under the AC field.
In Sec.~\ref{chap:noise_int}, the frequency dependence of photon-assisted current noises is discussed in detail.
We show that the vertex corrections are sensitive to the dynamics induced by the AC field.
The conclusions are given in Sec.~\ref{chap:conclusion}.

\section{FORMULATION}

\label{chap:formulation}

In this section, we describe our model and the general formulation for time-dependent transport of interacting electrons.~\cite{ding2013finite}
According to this scheme, the gauge invariance of transport quantities is guaranteed by self-consistently determined vertex functions.~\cite{PhysRev.124.287,PhysRev.127.1391}
We show the explicit expressions of the vertex functions within the SCHF approximation, and introduce Floquet-GFs.~\cite{tsuji2008correlated}

\subsection{Model}

We consider the single-impurity Anderson model with a time-dependent external field in order to describe the general properties of photon-assisted transport through a QD system. 
The model Hamiltonian is given as
\begin{align}
H = & \sum_{\sigma} (\epsilon_d + eA_{0\sigma}(t)) 
\hat{d}^{\dagger}_\sigma \hat{d}_\sigma, \nonumber \\
& + \sum_{\alpha = L, R}\sum_{{\bm k},\sigma} (\epsilon_{\bm k} + e v_{\alpha\sigma}(t)) 
\hat{c}^{\dagger}_{\alpha{\bm k}\sigma} \hat{c}_{\alpha{\bm k}\sigma}, \nonumber \\
& + \sum_{\alpha = L, R} \sum_{{\bm k},\sigma} \left(
t_{\alpha} e^{ieA_{\alpha\sigma}(t)} \hat{d}^{\dagger}_\sigma \hat{c}_{\alpha{\bm k}\sigma} 
+ {\rm h.c.} \right), \nonumber \\
\label{eq:Model Hamiltonian}
& + U \hat{n}_{\uparrow} \hat{n}_{\downarrow}.
\end{align}
Here, $\hat{d}_{\sigma}$ and $\hat{c}_{\alpha{\bm k}\sigma}$ are the electron annihilation operators of a single-level QD and reservoirs $\alpha$ ($=L,R$), respectively.
The electron spin is denoted by $\sigma$ $(=\uparrow, \downarrow)$, and the momentum of electrons in the reservoirs is denoted by ${\bm k}$.
The first term of the Hamiltonian describes an isolated QD with an energy level, $\epsilon_d$, under a gate voltage, $A_{0\sigma}(t)$,
whereas the second term represents noninteracting electron reservoirs with scalar potentials, $v_{\alpha\sigma}(t)$.
The third term describes electron tunneling between the QD and the leads,
where hopping amplitudes and vector potentials are denoted by $t_{\alpha}$ and $A_{\alpha \sigma}(t)$, respectively.~\cite{footnote1}
All the external fields, $A_{0\sigma}(t)$, $A_{\alpha \sigma}(t)$, and $v_{\alpha \sigma}(t)$, are assumed to be classical variables.
The Coulomb interaction in the QD is included in the last term. 
In this paper, we set the speed of light, $c$, and the Dirac constant, $\hbar$, to unity, and use the electron charge, $e (=-\left| e\right|<0)$.

\begin{figure}[t]
\centering
\includegraphics*[width=0.9\hsize]{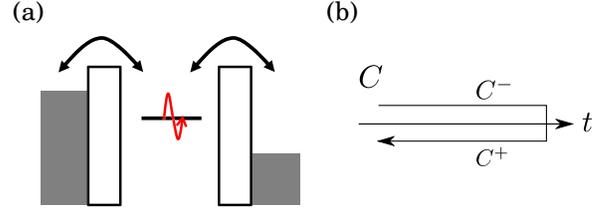}
\caption{\label{fig:System_Keldysh}
(a) Schematic picture of the system.
The energy level of the QD is sinusoidally modulated by the AC field.
(b) The Keldysh contour, $C$.
The upper (lower) branch is denoted by $C^{-(+)}$.
}
\end{figure}

In the actual calculations, we consider the case where the energy level of the QD is modulated from the symmetric point ($\epsilon_{d\sigma}=-U/2$) by an AC field; $A_{0\sigma}(t)=\epsilon_1 \sin(\Omega t)$ and $A_{L\sigma}(t)=A_{R\sigma}(t)=0$ [see Fig.~\ref{fig:System_Keldysh}(a)].
The bias voltage is assumed to be stationary and applied symmetrically; $v_{L\sigma}(t)=-v_{R\sigma}(t)=V/2$.
The chemical potentials of the leads are taken to be zero.
We are not concerned with spin-dependent transport in this paper.

We study nonequilibrium electron transport by implementing the Keldysh path integral formalism.~\cite{chou1985equilibrium,kamenev2011field} 
All the quantities are defined on the Keldysh contour, $C$, 
which is composed of the forward-time path, $C^-$, and the backward-time path, $C^+$ [see Fig.~\ref{fig:System_Keldysh}(b)].
The argument on the Keldysh contour is denoted by the Greek alphabet $\tau$.
We use the external fields defined on the Keldysh contour, $A_{0\sigma}(\tau)$, $A_{\alpha \sigma}(\tau)$, and $v_{\alpha \sigma}(\tau)$, which are allowed to have different values on the different branches $C^-$ and $C^+$.
The partition function, $Z[A]$, is written as
\begin{align}
& Z[A] = \int {\cal D}[\bar{d}d] e^{i(S_0 + S_{U})}, \\
& S_0 \equiv \int d\tau d\tau' \sum_{\sigma} \bar{d}_{\sigma}(\tau) G^{-1}_{0d\sigma}(\tau,\tau') d_{\sigma}(\tau'), \\
& S_{U} \equiv -\int d\tau Un_{d\uparrow}(\tau)n_{d\downarrow}(\tau),
\end{align}
where $d_{\sigma}$ and $\bar{d}_{\sigma}$ are the Grassmann fields of electrons in the QD.
The partition function, $Z[A]$, is a functional of the external fields, $A_{0\sigma}(\tau)$, $A_{\alpha\sigma}(\tau)$, and $v_{\alpha \sigma}(\tau)$, which are symbolically expressed as the argument, $A$.
$G_{0d \sigma}$ is the unperturbed GF of the dot electron, which includes the effect of hybridization between the QD and the leads:
\begin{align}
&G^{-1}_{0d\sigma}(\tau,\tau') \equiv g^{-1}_{d\sigma}(\tau,\tau')-\Sigma_{0\sigma}(\tau,\tau'),
\label{eq:def of the unperturbed GF} \\
&\Sigma_{0\sigma}(\tau,\tau') = \sum_{\alpha = L,R} \Sigma_{0 \alpha \sigma}(\tau,\tau'), \\\label{eq:def of tunneling self-energy}
&\Sigma_{0\alpha\sigma}(\tau,\tau') =\left|t_{\alpha}\right|^2 e^{ie(A_{\alpha\sigma}(\tau)-A_{\alpha\sigma}(\tau'))} \sum_{{\bm k}}g_{\alpha{\bm k}\sigma}(\tau,\tau').
\end{align}
Here, $g_{d\sigma}(\tau,\tau')$ and $g_{\alpha\bm{k}\sigma}(\tau,\tau')$ are the GFs of electrons for an isolated dot without a Coulomb interaction ($U=0$) and those of isolated leads, respectively.
Nonperturbative effects of the external fields are included in these unperturbed GFs.

The doubled degrees of freedom of the external gauge fields, $A^{\pm}_{0(\alpha)\sigma}(t)$, can describe both the time-evolution and the statistical correlation.
To preserve the normalization and the causality structure of the partition function,~\cite{kamenev2011field} the gauge fields, $A^{-}_{0(\alpha)\sigma}(t)$ and $A^{+}_{0(\alpha)\sigma}(t)$, must be equated to one another in the last step of calculations; $A^{-}_{0(\alpha)\sigma}(t)=A^{+}_{0(\alpha)\sigma}(t)=A_{0(\alpha)\sigma}(t)$.
Note that the physical external fields, $A_{0(\alpha)\sigma}(t)$, are left as finite quantities so that nonperturbative effects of the time-dependent field can be discussed.

\subsection{Formulation of time-dependent transport}

A systematic formulation of transport under time-dependent fields can be developed using two-particle-irreducible (2PI) formalism.~\cite{PhysRev.118.1417,PhysRevD.10.2428}
The 2PI effective action for the present system~\cite{footnote2} is written as
\begin{align}
\label{eq:CJT rule}
\Gamma[G;A]
= & - i \mathrm{Sp} \left[ \ln \left( \left. G_{0d\sigma} \right|_{A=0}G^{-1}_{d\sigma} \right) \right] \nonumber \\
&\hspace{20pt}- i \mathrm{Sp}\left[ G_{d\sigma} G^{-1}_{0d\sigma} -1 \right]  + \Gamma_2[G;A],
\end{align}
where $G$ is an abbreviation of the nonequilibrium GF $G_{d\sigma}(\tau,\tau')$.
The various components of the GF on the real time axis are summarized in Appendix~\ref{app:GFs}.
The symbol $\mathrm{Sp}$ denotes the summation of all the internal indices: the integration on the Keldysh contour and the summation of the spin.
The functional $\Gamma_2$ is the sum of the 2PI vacuum diagrams with the internal lines set to $G$, and corresponds to the Luttinger-Ward functional.~\cite{PhysRev.118.1417}

The Dyson equation can be derived by differentiating Eq.~(\ref{eq:CJT rule}) with respect to the propagator and setting the source fields to zero:
\begin{equation}
\label{eq:Dyson CTP}
G^{-1}_{d\sigma}(\tau,\tau') = G^{-1}_{0d\sigma}(\tau,\tau') -\Sigma_{U\sigma}(\tau,\tau'), \\
\end{equation}
where the self-energy is defined as
\begin{equation}
\label{eq:def of SE}
\Sigma_{U\sigma}(\tau,\tau') \equiv -i \frac{\delta \Gamma_2[G;A]}{\delta G_{d\sigma}(\tau',\tau)}.
\end{equation}

Current noises can be concisely written in terms of the current vertex functions
\begin{align}
\label{eq:def of the current vertex}
&\Gamma_{\sigma \alpha'\sigma'}(\tau_1,\tau_2;\tau') \nonumber \\
&\equiv \delta_{\sigma\sigma'} \Gamma_{0\alpha'\sigma'}(\tau_1,\tau_2;\tau')
+\Gamma_{U\sigma\alpha'\sigma'}(\tau_1,\tau_2;\tau'),
\end{align}
where the bare and dressed parts of the current vertex function are defined as
\begin{align}
\label{eq:def of the unperturbed current vertex}
&e\Gamma_{0\alpha'\sigma'}(\tau_1,\tau_2;\tau') 
\equiv - \frac{\delta G^{-1}_{0d\sigma'}(\tau_1,\tau_2)}{\delta A_{\alpha'\sigma'}(\tau')}, \\
\label{eq:def of the perturbed current vertex}
&e\Gamma_{U\sigma\alpha'\sigma'}(\tau_1,\tau_2;\tau') 
\equiv  \frac{\delta \Sigma_{U\sigma}(\tau_1,\tau_2)}{\delta A_{\alpha'\sigma'}(\tau')},
\end{align}
respectively.
The bare vertex function is obtained as 
\begin{align}
\label{eq:unp. vert. func. and tun. SE}
&\Gamma_{0\alpha'\sigma'}(\tau_1,\tau_2;\tau') = -i \left[ \delta(\tau',\tau_2)-\delta(\tau_1,\tau') \right] \nonumber \\
& \hspace{110pt} \times \Sigma_{0\alpha'\sigma'} (\tau_1,\tau_2),
\end{align}
whereas the dressed vertex function is related to the self-energy by the relation
\begin{align}
\label{eq:identity}
&\frac{\delta G_{d\sigma}(\tau_1,\tau_2)}{\delta A_{\alpha'\sigma'}(\tau')}
=e\int d\tau_3d\tau_4 G_{d\sigma}(\tau_1,\tau_3) \Gamma_{\sigma \alpha'\sigma'}(\tau_3,\tau_4;\tau') \nonumber \\
&\hspace{130pt} \times G_{d\sigma}(\tau_4,\tau_2).
\end{align}
The gauge invariance of the transport quantities is guaranteed by the vertex functions self-consistently determined using these relations.~\cite{PhysRev.124.287,PhysRev.127.1391}

The current-current correlation function on the Keldysh contour is defined by the functional derivative of the 2PI effective action as
\begin{align}
\label{eq:def of current-current corr.}
D_{\alpha\sigma\alpha'\sigma'}(\tau,\tau') &\equiv -i \left. \frac{\delta^2 \Gamma[G;A]}{ \delta A_{\alpha\sigma}(\tau)  \delta A_{\alpha'\sigma'}(\tau')} \right|_{A^s=K=0} \nonumber \\
&=e^2 \left[ \langle  j_{\alpha\sigma}(\tau)j_{\alpha'\sigma'}(\tau') \rangle  \right.\nonumber \\
&\hspace{40pt} \left. - \langle j_{\alpha\sigma}(\tau) \rangle \langle j_{\alpha'\sigma'}(\tau') \rangle  \right].
\end{align}
The current-current correlation function is written as the sum of the two terms:
\begin{equation}
\label{c-c corr. division}
D_{\alpha\sigma\alpha'\sigma'}(\tau,\tau')=D_{0\alpha\alpha'\sigma}(\tau,\tau')\delta_{\sigma\sigma'}+D_{U\alpha\sigma\alpha'\sigma'}(\tau,\tau'),
\end{equation}
where
\begin{widetext}
\begin{align}
D_{0\alpha\alpha'\sigma}(\tau,\tau')
\equiv & -i e^2\delta_{\alpha\alpha'} \int d\tau_1   \left[ G_{d\sigma}(\tau',\tau_1)\Gamma_{0\alpha\sigma}(\tau_1,\tau';\tau)- \Gamma_{0\alpha\sigma}(\tau',\tau_1;\tau)G_{d\sigma}(\tau_1,\tau')  \right] \nonumber \\
\label{eq:current-current unp.}
&  +e^2  \int d\tau_1d\tau_2d\tau_3d\tau_4 G_{d\sigma}(\tau_1,\tau_2)\Gamma_{0\alpha'\sigma}(\tau_2,\tau_3;\tau') G_{d\sigma}(\tau_3,\tau_4)\Gamma_{0\alpha\sigma}(\tau_4,\tau_1;\tau),
\end{align}
\begin{align}
D_{U\alpha\sigma\alpha'\sigma'}(\tau,\tau')
&\equiv e^2\int d\tau_1d\tau_2d\tau_3d\tau_4 G_{d\sigma}(\tau_1,\tau_2)\Gamma_{U\sigma\alpha'\sigma'}(\tau_2,\tau_3;\tau') G_{d\sigma}(\tau_3,\tau_4)\Gamma_{0\alpha\sigma}(\tau_4,\tau_1;\tau).
\end{align}
\end{widetext}
The former is called the bare part of the current-current correlation and the latter is its vertex correction.
These are the formal expressions of the current noises, which have been obtained by Ding and Dong~\cite{ding2013finite} using the full counting statistics.

The symmetrized current noises are defined as
\begin{equation}
S_{\alpha\sigma\alpha'\sigma'}(t,t')  \equiv S_{0\alpha\alpha'\sigma}(t,t')\delta_{\sigma\sigma'}+S_{U\alpha\sigma\alpha'\sigma'}(t,t'),
\end{equation}
where
\begin{align}
\label{eq:sym. cur. noise unp.}
&S_{0\alpha\alpha'\sigma}(t,t') \equiv D^{-+}_{0\alpha\alpha'\sigma}(t,t')+D^{+-}_{0\alpha\alpha'\sigma}(t,t'), \\
\label{eq:sym. cur. noise VC}
&S_{U\alpha\sigma\alpha'\sigma'}(t,t') \equiv D^{-+}_{U\alpha\sigma\alpha'\sigma'}(t,t')+D^{+-}_{U\alpha\sigma\alpha'\sigma'}(t,t').
\end{align}
The vertex corrections, which cannot be expressed solely in terms of the one-body quantity, are essential for satisfying the gauge invariance of the current noises.

\subsection{Vertex functions within the SCHF approximation}
\label{subsec:ver}

In this paper, we use the SCHF approximation.~\cite{PhysRevB.46.7061,ding2013finite}
The self-energy is given by the Hartree term as
\begin{equation}
\label{eq:SE}
\Sigma_{U\sigma}(\tau_1,\tau_2)=-iU G_{d\bar{\sigma}}(\tau_1,\tau_2)\delta(\tau_1,\tau_2),
\end{equation}
where $\delta(\tau,\tau')$ is the Dirac delta function on the Keldysh contour.
The dressed vertex functions can be written as
\begin{equation}
\Gamma_{U\sigma\alpha'\sigma'}(\tau_1,\tau_2;\tau') \equiv \tilde{\Gamma}_{U\sigma\alpha'\sigma'}(\tau_1,\tau')\delta(\tau_1,\tau_2), 
\end{equation}
because the incoming and outgoing fermion lines meet at the same vertex in the SCHF approximation.
We define the bare parts of the density-density, density-current, and current-density correlation functions as
\begin{align}
\label{eq:nnbare}
&\chi_{0nn\sigma}(\tau,\tau')
\equiv -i G_{d\sigma}(\tau,\tau')G_{d\sigma}(\tau',\tau),\\ 
\label{eq:ncbare}
&\chi_{0n\alpha\sigma}(\tau,\tau')
\equiv i\int d\tau_1 d\tau_2 G_{d\sigma}(\tau,\tau_1) \nonumber \\
&\hspace{85pt} \times \Gamma_{0\alpha\sigma}(\tau_1,\tau_2;\tau') G_{d\sigma}(\tau_2,\tau), \\
\label{eq:cnbare}
&\chi_{0\alpha n\sigma}(\tau,\tau') \equiv \chi_{0n\alpha \sigma}(\tau',\tau).
\end{align}
Furthermore, we define the polarization function, $M_{\sigma}(\tau,\tau')$, which satisfies the integral equation as
\begin{align}
\label{eq:def of M}
\left( \left[1 -U^2 \chi_{0nn\bar{\sigma}}\chi_{0nn\sigma} \right] M_{\sigma} \right) (\tau,\tau') =\delta(\tau,\tau'),
\end{align}
where $\left( AB\right)(\tau,\tau')\equiv \int d\tau_1 A(\tau,\tau_1)B(\tau_1,\tau')$.
Using these functions, the dressed vertex functions are written as
\begin{align}
&\tilde{\Gamma}_{U\bar{\sigma}\alpha\sigma}(\tau,\tau')=-U \left( M_{\sigma}\chi_{0n\alpha\sigma} \right) (\tau,\tau'), \\
&\tilde{\Gamma}_{U\sigma\alpha\sigma}(\tau,\tau')=-U^2 \left( M_{\sigma} \chi_{0nn\bar{\sigma}} \chi_{0n\alpha\sigma} \right) (\tau,\tau').
\end{align}

\subsection{Floquet-GFs}
\label{app:FGF}

We define the Fourier transformation of the GFs with respect to the relative time, $t_r \equiv t-t'$, as
\begin{align}
G^{\nu\nu'}_{d\sigma}(\omega,T) &\equiv \int_{-\infty}^{\infty} dt_r e^{i\omega t_r} G^{\nu\nu'}_{d\sigma}(T+t_r/2, T-t_r/2),
\end{align}
where $\nu$ and $\nu'$ are the Keldysh indices and $T\equiv (t+t')/2$ is the average time.
The GFs of systems driven by a periodically oscillating field are invariant under discrete-time translations, $t \rightarrow t+T_p$, where $T_p\equiv 2\pi / \Omega$ is the period of the AC field.
Then, the Wigner representation of the GF is defined as
\begin{align}
\left( G^{\nu\nu'}_{d\sigma}\right)_n(\omega) &\equiv \frac{1}{T_p}\int _{-T_p/2}^{T_p/2}dT e^{in\Omega T}G^{\nu\nu'}_{d\sigma}(\omega,T).
\end{align}
The Wigner representation is suitable for gaining physical insights because of its clear interpretation.
In particular, the zeroth mode of the Wigner representation $(G^{\nu\nu'}_{d\sigma})_0(\omega)$ corresponds to the GF averaged over one period of the AC field.

The Fourier indices can be efficiently handled if the GF is transformed into the Floquet representation~\cite{tsuji2008correlated} as
\begin{eqnarray}
\label{eq:def of FGF}
  \left(\bm{G}^{\nu\nu'}_{d\sigma}\right)_{mn}(\omega)&\equiv& \left(G^{\nu\nu'}_{d\sigma}\right)_{m-n}\left(\omega +\frac{m+n}{2}\Omega \right),
\end{eqnarray}
where the frequency $\omega$ is folded into the first time-Brillouin zone, i.e., $-\Omega/2 \leq \omega < \Omega/2$.
In this paper, we use bold letters for matrices in the Floquet representation.
The Wigner and Floquet representations of other quantities (self-energy, current noises, etc.) can be defined in the same way.

\begin{figure}[t]
\centering
\includegraphics*[width=0.7\hsize]{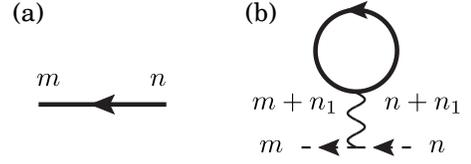}
\caption{\label{fig:FGFs}
Diagrammatic representation of (a) Floquet-GF and (b) Hartree term.
The Floquet indices are allocated at each endpoint of the GFs.
}
\end{figure}

Feynman rules in a driven system are analogous to those in a time-translationally invariant system, except for additional Floquet indices.
The Floquet-GF $({\bm G}^{\nu\nu'}_{d\sigma})_{mn}(\omega)$ is expressed by a propagator with Floquet indices $n$ and $m$ at the initial and terminal points, respectively [Fig.~\ref{fig:FGFs}(a)].
The internal Floquet indices are summed up at all the vertices of the diagram.
For instance, the Hartree term [Fig.~\ref{fig:FGFs}(b)] includes the summation of the additional index $n_1$.

Various equations can be expressed in a simple matrix form using the Floquet representation.
In particular, the exact equations of the retarded and lesser Floquet-GF of the QD can be derived from Eq.~(\ref{eq:Dyson CTP}) as
\begin{align}
\label{eq:Dyson}
{\bm G}^r_{d\sigma}(\omega)&={\bm G}^r_{0d\sigma}(\omega)+{\bm G}^r_{0d\sigma}(\omega){\bm \Sigma}^r_{U\sigma}(\omega)\bm{G}^r_{d\sigma}(\omega),\\
\label{eq:Keldysh}
{\bm G}^{-+}_{d\sigma}(\omega) &=  {\bm G}^r_{d\sigma}(\omega) \left({\bm \Sigma}^{-+}_{0\sigma}(\omega)+{\bm \Sigma}^{-+}_{U\sigma}(\omega)\right){\bm G}^a_{d\sigma}(\omega).
\end{align}
Eqs.~(\ref{eq:Dyson}) and (\ref{eq:Keldysh}) are called the Dyson and Keldysh equation, respectively.

The relations of the Floquet-GFs are inherited from the GFs defined on the real-time axis.
The advanced Floquet-GF is determined by the relation ${\bm G}^a_{d\sigma}(\omega)=\left( {\bm G}^r_{d\sigma}\right)^{\dagger} (\omega) $.
The lesser and greater Floquet-GFs have the relations ${\bm G}^{-+}_{d\sigma}(\omega)=-\left( {\bm G}^{-+}_{d\sigma}\right)^{\dagger} (\omega) $ and ${\bm G}^{+-}_{d\sigma}(\omega)=-\left( {\bm G}^{+-}_{d\sigma} \right)^{\dagger}(\omega) $, respectively.

\section{Generalized distribution function and spectral function}
\label{chap:PAT}

In this section, we define the generalized distribution function to understand the characteristic properties of driven systems.
In addition, we calculate the spectral function within the SCHF approximation. 
Throughout this paper, we assume $\Delta_L=\Delta_R=1$, $\epsilon_d=-U/2$ (the particle-hole symmetric condition), $\beta=20$, and $\Omega=5$ in all numerical calculations.

\subsection{Generalized distribution function}
\label{sec:unp FGFs}

The tunneling self-energy (\ref{eq:def of tunneling self-energy}) describes nonequilibrium electron tunneling between the leads and the QD.
The tunneling self-energy is calculated as~\cite{tsuji2008correlated} 
\begin{align}
\label{eq:ret. tun. SE}
{\bm \Sigma}^{r}_{0\alpha\sigma}(\omega)&=-i\frac{\Delta_\alpha}{2}{\bm 1}, \\
{\bm \Sigma}^{-+}_{0\alpha\sigma}(\omega)&=i\Delta_{\alpha}{\bm J}^{\dagger}
\tilde{{\bm f}}_{\alpha}(\omega){\bm J}, \label{eq:les. tun. SE} 
\end{align}
in the wide band limit. 
The line width is defined as $\Delta_{\alpha}\equiv 2\pi\left|t_{\alpha}\right|^2\rho_0$, where $\rho_0$ is the DOS of a conduction electron at the Fermi energy.
The unitary matrix, ${\bm J}$, is defined using the Bessel function, $J_m(\epsilon_1/\Omega)$, as $\left({\bm J}\right)_{mn}\equiv i^{m-n}J_{m-n}(\epsilon_1/\Omega)$, and ${\bm 1}$ is a unit matrix.
The diagonal matrix, $\tilde{{\bm f}}_{\alpha}$, is defined as $(\tilde{{\bm f}}_{\alpha})_{mn}(\omega)\equiv f\left(\omega+m\Omega -ev_{\alpha\sigma}+\epsilon_d\right) \delta_{mn}$, where $f(\omega) = (e^{\beta \omega}+1)^{-1}$.

The lesser component of the tunneling self-energy can be rewritten as
\begin{align}
{\bm \Sigma}^{-+}_{0\alpha\sigma}(\omega) &=i\Delta_\alpha {\bm f}_{\alpha}(\omega), \\
{\bm f}_{\alpha}(\omega) &\equiv {\bm J}^{\dagger}\tilde{{\bm f}}_{\alpha}(\omega){\bm J},
\end{align}
by introducing the generalized nonequilibrium distribution function, ${\bm f}_{\alpha}(\omega)$.
We note that ${\bm f}_{\alpha}(\omega)$ coincides with $\tilde{{\bm f}}_{\alpha}(\omega)$ in the absence of the AC field ($\epsilon_1=0$) because ${\bm J}$ becomes a unit matrix.

\begin{figure*}[t]
\centering
\includegraphics*[width=0.8\hsize]{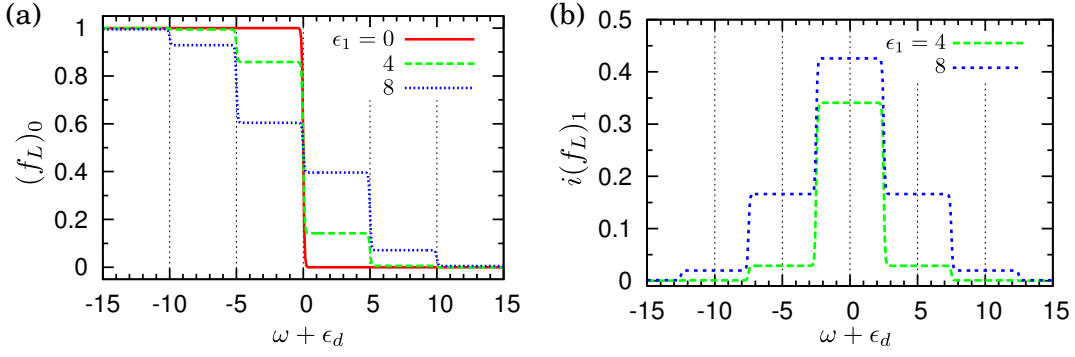}
\caption{\label{fig:GDF}
(Color online) (a) The zeroth mode of the generalized distribution function, $(f_{L})_{0}(\omega)$, for $\epsilon_1=0$, $4$, and $8$.
(b) The first mode of the generalized distribution function, $i(f_{L})_{1}(\omega)$.
The amplitude of the AC field is taken to be $\epsilon_1=4$ and $8$.
Parameters are as follows: $\Delta_L=\Delta_R=1$, $\epsilon_d=-U/2=-1$, $\beta=20$, $V=0$, and $\Omega=5$.}
\end{figure*}

In this paper, the effect of the AC field is fully included in the generalized distribution function, ${\bm f}_{\alpha}(\omega)$, while the unperturbed dot GFs are not modified.~\cite{footnote3}
Features of ${\bm f}_{\alpha}(\omega)$ can be clearly seen in the Wigner representation.
The zeroth mode of the generalized distribution function, $(f_{\alpha})_0(\omega)$, is written as the weighted sum of the Fermi distribution function:
\begin{align}
(f_\alpha)_0(\omega)
= \sum J_m\left( \frac{\epsilon_1}{\Omega} \right)^2 f(\omega + m\Omega - ev_{\alpha} + \epsilon_d).
\end{align}
Figure~\ref{fig:GDF}(a) displays $(f_L)_0(\omega)$ at $\epsilon_d=-U/2=-1$ and $V=0$ for $\epsilon_1=0$, $4$, and $8$.
The width of each step is equal to the external driving frequency, $\Omega$, and its height at $\omega=n\Omega$ ($n\in{\bm Z}$) is the square of the Bessel function, $\left| J_{n}(\frac{\epsilon_1}{\Omega})\right|^2$.
We note that the multi-step structure of the generalized distribution function originates from the coherent nature of electrons under the AC field.
The first mode of the generalized distribution function, $i\left(f_L\right)_{1}(\omega) (=-i\left(f_L\right)_{-1}(\omega))$, is shown in Fig.~\ref{fig:GDF}(b) for $\epsilon_1=4$ and $8$, whose contribution is not small in comparison with the zeroth mode.

The Floquet-GFs for $U=0$, which are hereafter called the unperturbed Floquet-GFs, are calculated as 
\begin{eqnarray}
\label{eq:unperturbed}
{\bm G}^r_{0d\sigma} (\omega)&=&\frac{1}{\omega+m\Omega+i\Delta/2}{\bm 1},\nonumber \\
{\bm G}^{-+}_{0d\sigma} (\omega)&=&i\Delta {\bm G}^{r}_{0d\sigma} (\omega){\bm f}(\omega){\bm G}^{a}_{0d\sigma}(\omega) ,
\end{eqnarray}
with $\Delta\equiv\Delta_L+\Delta_R$ and
\begin{equation}
 {\bm f}(\omega)\equiv\frac{\Delta_L{\bm f_L}(\omega)+\Delta_R{\bm f_R}(\omega)}{\Delta_L+\Delta_R}.
\end{equation}
By utilizing the Floquet-GFs, the lesser component is written in a pseudo-equilibrium form~\cite{haug2008quantum} even with an AC field.

\subsection{Spectral function}
\label{subsec:DOS}

\begin{figure}[t]
\centering
\includegraphics*[width=0.8\hsize]{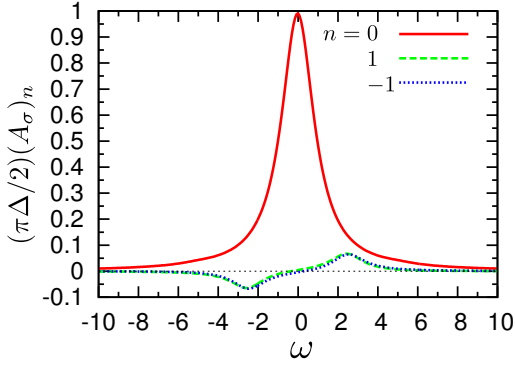}
\caption{\label{fig:DOS}
The three different modes of the spectral function: $(A_{\sigma})_0(\omega)$, $(A_{\sigma})_1(\omega)$, and $(A_{\sigma})_{-1}(\omega)$.
Parameters are as follows: $\Delta_L=\Delta_R=1$, $\epsilon_d=-U/2=-1.5$, $\beta=20$, $V=0$, $\epsilon_1=8$, and $\Omega=5$.}
\end{figure}

The self-energy within the SCHF approximation is given by the Hartree diagrams [Fig.~\ref{fig:FGFs}(b)];
\begin{align}
\label{eq:SE_r}
&(\bm{\Sigma}_{U\sigma}^{r})_{mn}(\omega)=-iU\sum_{n_1}\int_{-\frac{\Omega}{2}}^{\frac{\Omega}{2}}\frac{d\omega_1}{2\pi}(\bm{G}^{-+}_{d\bar{\sigma}})_{{n_1}+m-n,{n_1}}(\omega_1), \\
\label{eq:SE_mp}
&(\bm{\Sigma}_{U\sigma}^{-+})_{mn}(\omega)=0.
\end{align}
We note that the integrand of the retarded self-energy is the full Floquet-GF.
Using these expressions, the Dyson equation (\ref{eq:Dyson}) and the Keldysh equation (\ref{eq:Keldysh}) are simplified to
\begin{align}
&{\bm G}^r_{d\sigma}(\omega)= \left[ {\bm 1} - {\bm G}^r_{0d\sigma}(\omega){\bm \Sigma}_{U\sigma}(\omega) \right]^{-1}{\bm G}^r_{0d\sigma}(\omega), \\
\label{eq:Lesser FGF}
&{\bm G}^{-+}_{d\sigma}(\omega)=i\Delta {\bm G}^{r}_{d\sigma}(\omega) {\bm f}(\omega) {\bm G}^{a}_{d\sigma}(\omega),
\end{align}
respectively.
The Floquet-GFs are determined by solving Eqs.~(\ref{eq:SE_r})-(\ref{eq:Lesser FGF}) self-consistently.

The Wigner representation of the spectral function is defined as
\begin{equation}
(A_{\sigma})_n(\omega)\equiv -\frac{1}{\pi}\  {\rm Im}\left(G^r_{d\sigma}\right)_n(\omega),
\end{equation}
where $(G^r_{d\sigma})_n$ is the Wigner representation of the retarded GF.
Figure~\ref{fig:DOS} shows $(A_{\sigma})_0(\omega)$, $(A_{\sigma})_1(\omega)$, and $(A_{\sigma})_{-1}(\omega)$ for $\epsilon_d=-U/2=-1.5$, $V=0$, and $\epsilon_1=8$.
The zeroth mode of the spectral function, $(A_{\sigma})_0(\omega)$, has a Lorentzian spectral function of the QD with a level broadening $\Delta$.
The different oscillating modes of the spectral function $(A_{\sigma})_{\pm1}(\omega)$ have a peak (dip) around $\omega=\pm \Omega/2$ because the imaginary part of the self-energy becomes finite for the off-diagonal components.
The non-negativity of the spectral function does {\it not} necessarily hold true in nonequilibrium systems, while a spectral momentum sum rule,~\cite{PhysRevB.73.075108}
\begin{equation}
\label{eq:0th SMSR }
\int d\omega  \left(A_{\sigma}\right)_n(\omega)=\delta_{n0},
\end{equation}
can be confirmed by our result.

\section{Photon-assisted current noise at zero frequency}

\label{chap:noise_non}

In this section, we study the zero-frequency current noise under the AC field.
We introduce an effective temperature to characterize the effects of photon absorption and emission processes on the current noises.

The photon-assisted current noises in the Floquet representation are given by the sum of the bare part and the vertex correction as
\begin{align}
{\bm S}_{\alpha \sigma \alpha' \sigma'}(\omega) =
{\bm S}_{0 \alpha \alpha'\sigma}(\omega) \delta_{\sigma \sigma'} + {\bm S}_{U\alpha\sigma\alpha'\sigma'}(\omega).
\end{align}
The explicit expression of the bare part can be obtained by straightforward calculation from Eq.~(\ref{eq:current-current unp.}) as
\begin{widetext}
\begin{align}
\label{eq:unp. f.f. current noise}
{\bm S}_{0\alpha\alpha'\sigma}(\omega) /e^2 
&=\delta_{\alpha\alpha'} \left[  \left(  {\bm G}^{-+}_{d\sigma} \circ {\bm \Sigma}^{+-}_{0\alpha\sigma} \right)(\omega) + \left(  {\bm \Sigma}^{-+}_{0\alpha\sigma} \circ {\bm G}^{+-}_{d\sigma} \right)(\omega)  \left(  {\bm G}^{+-}_{d\sigma} \circ {\bm \Sigma}^{-+}_{0\alpha\sigma}   \right)(\omega) + \left(  {\bm \Sigma}^{+-}_{0\alpha\sigma} \circ {\bm G}^{-+}_{d\sigma} \right)(\omega) \right] \nonumber \\
&\hspace{10pt} - \left[ \left(  {\bm G}^{r}_{d\sigma} {\bm \Sigma}^{-+}_{0\alpha'\sigma} \circ  {\bm \Sigma}^{+-}_{0\alpha'\sigma} {\bm G}^{a}_{d\sigma} \right)(\omega)  + \left( {\bm G}^{r}_{d\sigma}  {\bm \Sigma}^{+-}_{0\alpha'\sigma} \circ {\bm \Sigma}^{-+}_{0\alpha'\sigma} {\bm G}^{a}_{d\sigma} \right)(\omega) \right. \nonumber \\
&\hspace{60pt}  +  \left( {\bm \Sigma}^{+-}_{0\alpha\sigma} {\bm G}^{a}_{d\sigma} \circ  {\bm G}^{r}_{d\sigma} {\bm \Sigma}^{-+}_{0\alpha\sigma} \right)(\omega) +\left( {\bm \Sigma}^{-+}_{0\alpha\sigma} {\bm G}^{a}_{d\sigma} \circ {\bm G}^{r}_{d\sigma} {\bm \Sigma}^{+-}_{0\alpha\sigma}  \right)(\omega) \nonumber \\
&\hspace{60pt} \left. -2 \Delta_{\alpha} \Delta_{\alpha'} \left(  {\bm G}^{r}_{d\sigma}{\bm f}_{\alpha'} - {\bm f}_{\alpha} {\bm G}^{a}_{d\sigma} + {\bm G}^{-+}_{d\sigma} \circ {\bm G}^{r}_{d\sigma}{\bm f}_{\alpha} - {\bm f}_{\alpha'} {\bm G}^{a}_{d\sigma} + {\bm G}^{-+}_{d\sigma} \right)(\omega)   \right] ,
\end{align}
where the circle denotes the convolution in the Floquet representation,
\begin{align}
\label{eq:bubble FGF}
 \left( {\bm A} \circ {\bm B} \right)_{mn}(\omega) \equiv \sum_{m_1,n_1} \int_{-\frac{\Omega}{2}}^{\frac{\Omega}{2}}\frac{d\omega_1}{2\pi} ({\bm A})_{m_1,n_1}(\omega_1) ({\bm B})_{-n+n_1-N_1,-m+m_1-N_1}(-\omega+\omega_1+N_1\Omega).
\end{align}
\end{widetext}
The integer $N_1$ is chosen so that the argument of the function ${\bm B}$ is reduced into the first time-Brillouin zone, i.e., $-\frac{\Omega}{2} \leq -\omega+\omega_1+N_1\Omega  <  \frac{\Omega}{2}$.

In the SCHF approximation, the vertex corrections to the photon-assisted current noises for the parallel and anti-parallel spins are calculated as
\begin{align}
&{\bm S}_{U\alpha\sigma\alpha'\sigma}(\omega) /e^2 \nonumber \\
&= iU^2 \left[ {\bm \chi}^r_{0\alpha n\sigma}(\omega){\bm M}^r_{\sigma}(\omega){\bm \chi}^{r}_{0nn\bar{\sigma}}(\omega) {\bm \chi}^{K}_{0n\alpha'\sigma}(\omega)\right. \nonumber\\
&\hspace{30pt} +{\bm \chi}^r_{0\alpha n\sigma}(\omega){\bm M}^r_{\sigma}(\omega) {\bm \chi}^{K}_{0nn\bar{\sigma}}(\omega)  {\bm \chi}^{a}_{0n\alpha'\sigma}(\omega) \nonumber\\
&\hspace{30pt} +{\bm \chi}^r_{0\alpha n\sigma}(\omega) {\bm M}^{K}_{\sigma}(\omega) {\bm \chi}^{a}_{0nn\bar{\sigma}}(\omega){\bm \chi}^{a}_{0n\alpha'\sigma}(\omega) \nonumber\\
&\hspace{30pt} +\left. {\bm \chi}^{K}_{0\alpha n\sigma}(\omega) {\bm M}^{a}_{\sigma}(\omega) {\bm \chi}^{a}_{0nn\bar{\sigma}}(\omega){\bm \chi}^{a}_{0n\alpha'\sigma}(\omega)\right],
\end{align}
and
\begin{align}
{\bm S}_{U\alpha\bar{\sigma}\alpha'\sigma}(\omega) /e^2 
&=iU \left[{\bm \chi}^r_{0\alpha n\bar{\sigma}}(\omega){\bm M}^r_{\sigma}(\omega){\bm \chi}^{K}_{0n\alpha'\sigma}(\omega)  \right. \nonumber \\
& \hspace{10pt}  +{\bm \chi}^r_{0\alpha n\bar{\sigma}}(\omega){\bm M}^{K}_{\sigma}(\omega){\bm \chi}^{a}_{0n\alpha'\sigma}(\omega) \nonumber \\
\label{eq:VC f.f. current noise}
& \hspace{10pt} \left. +{\bm \chi}^{K}_{0\alpha n\bar{\sigma}}(\omega){\bm M}^{a}_{\sigma}(\omega) {\bm \chi}^{a}_{0n\alpha'\sigma}(\omega) \right],
\end{align}
respectively.
The expressions of the vertex correction terms in the absence of the AC field have already been obtained for the zero-frequency noise~\cite{PhysRevB.46.7061} 
and for the finite-frequency noise.~\cite{ding2013finite}
The present results are straightforward extensions of these previous works in consideration of the finite AC field.

\begin{figure}[t]
\centering
\includegraphics*[width=0.9\hsize]{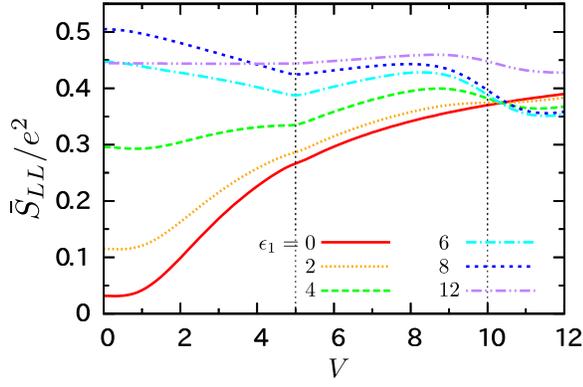}
\caption{\label{fig:photon-assisted noise}
(Color online)
Zero-frequency photon-assisted current noise for $\epsilon_1=0$, $2$, $4$, $6$, $8$, and $12$.
Parameters are as follows: $\Delta_L=\Delta_R=1$, $\epsilon_d=-U/2=-1$, $\beta=20$, and $\Omega=5$.}
\end{figure}

\begin{figure*}[t]
\centering
\includegraphics*[width=0.8\hsize]{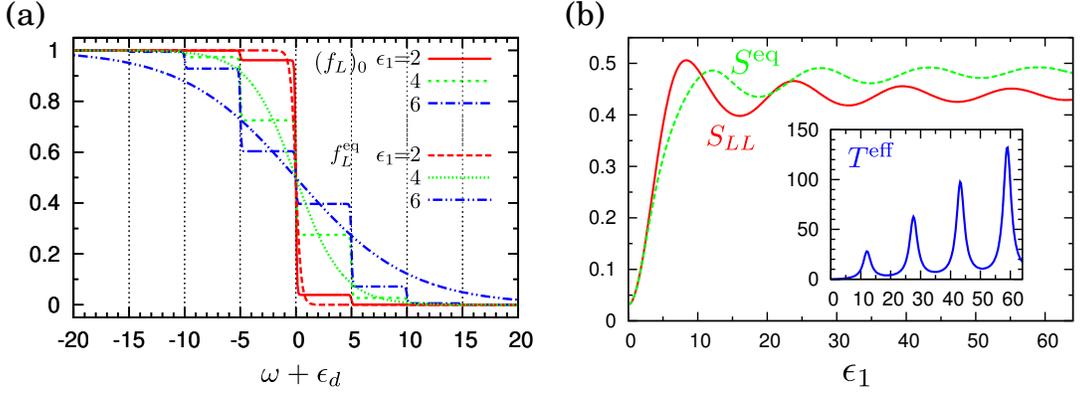}
\caption{\label{fig:Noise_eps}
(Color online)
(a) Staircase-structured generalized distribution function, $(f_L)_0$, compared with the smooth equilibrium Fermi distribution function, $f^{\rm eq}_{L}$, at $T=T_{\rm{eff}}$.
The external field amplitudes are for $\epsilon_1=2$ , $6$, and $8$.
(b) Comparison between the photon-assisted thermal noise and the equilibrium thermal noise evaluated at the corresponding $T_{\rm{eff}}$.
The dependence of the effective temperature, $T_{\rm{eff}}$, on the amplitude of the external AC field, $\epsilon_1$, is plotted in the inset.
Parameters are as follows: $\Delta_L=\Delta_R=1$, $\epsilon_d=-U/2=-1$, $\beta=20$, $V=0$, and $\Omega=5$.
}
\end{figure*}

Figure~\ref{fig:photon-assisted noise} shows the spin-averaged zero-frequency current noise, $\bar{S}_{LL}\equiv (1/2)\sum_{\sigma\sigma'}\left(S_{L\sigma L\sigma'}\right)_0(0)$, as a function of the bias voltage, $V$, for $\epsilon_d=-U/2=-1$.
We can see two significant features: (1) the singularities at the bias voltage corresponding to the driving frequency $\Omega$ and (2) the remarkable enhancement of the zero-bias current noise by the AC field.
The singularities of $\bar{S}_{LL}$ at $\Omega$ reflects the structures of the generalized distribution function [see Fig.~\ref{fig:GDF}(a)].
This feature is consistent with the previous work, which used the energy-independent scattering matrix.~\cite{lesovik1994noise}
The enhancement of the zero-bias noise by the AC field can be understood in terms of an effective temperature.

Roughly speaking, the multi-step structure of the generalized distribution function can be approximated by the Fermi distribution function with a modified temperature.
In this paper, we define an effective temperature by an extrapolated fluctuation dissipation relation:~\cite{reydellet2003quantum,caso2012defining}
\begin{equation}
\label{eq:T_eff}
T_{\rm{eff}} \equiv \frac{\bar{S}_{LL}}{4k_B \bar{G}},
\end{equation}
where $k_B$ is the Boltzmann constant.
The linear conductance is defined as $\bar{G}\equiv {\rm lim}_{V\rightarrow 0} (\bar{I}/V)$ by the time-averaged current
\begin{align}
\bar{I} = \frac{e\Delta_L \Delta_R}{\Delta_L + \Delta_R}\sum_{m} \int_{-\frac{\Omega}{2}}^{\frac{\Omega}{2}}\frac{d\omega_1}{2\pi} \left({\bm A}_{\sigma} ({\bm f}_L - {\bm f}_R)   \right)_{mm}(\omega_1).
\end{align}
We observe that the complete definition of the effective temperature has been open to discussion.~\cite{casas2003temperature}

In Fig.~\ref{fig:Noise_eps}(a), the generalized distribution function, $(f_L)_0$, is compared with the Fermi distribution function, $f^{\rm eq}_{L}$, with $T=T_{\rm eff}$ for $\epsilon_d=-U/2=-1$ and $V=0$.
The overall broadening of the multi-step structures is well-approximated by the Fermi distribution with the corresponding effective temperatures for $\epsilon_1=2$ and $6$, while $f^{\rm eq}_{L}$ no longer coincides with $(f_L)_0$ for $\epsilon_1=8$.

In Fig.~\ref{fig:Noise_eps}(b), we show $S_{LL}$ and the {\it equilibrium} thermal noise, $S^{\rm eq}$, with $T=T_{\rm eff}$ as a function of $\epsilon_1$ for $\epsilon_d=-U/2=-1$ and $V=0$. 
These quantities agree well with each other for $\epsilon_1<\Omega$, indicating that the effective temperature works well for the weak AC field.
For the strong AC field, $S_{LL}$ oscillates as a function of $\epsilon_1$ due to the coherent nature of electrons under the strong AC field.
This oscillation has already been obtained
in noninteracting electron systems.~\cite{gasse2013observation,hammer2011quantum}
The present result indicates that the quantum oscillation in photon-assisted current noises 
is robust against the Coulomb interaction.
The oscillation of $S^{\rm eq}$ can be explained by the dependence of the effective temperature, $T_{\rm eff}$, on $\epsilon_1$ [see the inset of Fig.~\ref{fig:Noise_eps}(b)].
$T_{\rm eff}$ is significantly enhanced at $\epsilon_1/\Omega = 2.40$, $5.52$, $8.65$, and $11.7$, which correspond to the zeros of the zero-order Bessel function, $J_0(\epsilon_1/\Omega)$.
The difference in the positions of the peaks and dips between $S_{LL}$ and $S^{\rm eq}$ implies the limit of the present definition of the effective temperature.

\section{Photon-assisted current noises at finite frequencies}
\label{chap:noise_int}

In this section, we study the frequency dependence of various correlation functions, polarization functions, and photon-assisted current noises employing the SCHF approximation.
The effects of the AC field on these quantities are discussed from the point of view of the effective temperature and photon absorption and emission.

\subsection{Correlation functions}
\label{sec:resp. func.}

\begin{figure*}[t]
\centering
\includegraphics*[width=0.8\hsize]{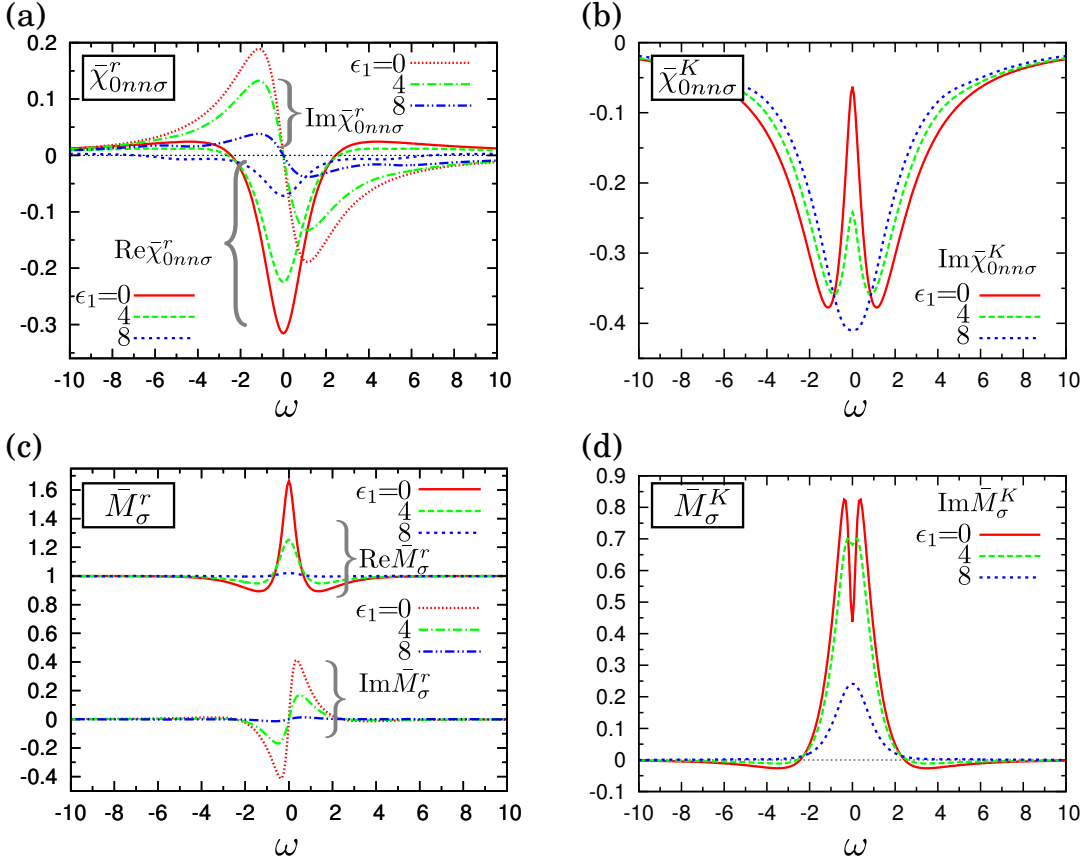}
\caption{\label{fig:density_density}
(Color online)
The retarded and the Keldysh components of the charge-charge correlation function [(a) and (b)] and the polarization function [(c) and (d)] for $\epsilon_1=0$, $4$, and $8$.
Parameters are as follows: $\Delta_L=\Delta_R=1$, $\epsilon_d=-U/2=-1$, $\beta=20$, $V=0$, and $\Omega=5$.
}
\end{figure*}

The vertex corrections to the photon-assisted current noises are written in terms of the bare correlation functions and the polarization function.
We discuss their properties under the AC field in this subsection.

In the Floquet representation, the retarded and lesser components of the charge-charge correlation function Eq.~(\ref{eq:nnbare}) are written as
\begin{align}
&{\bm \chi}^r_{0nn\sigma}(\omega)=-i \left[ ({\bm G}^r_{d\sigma} \circ {\bm G}^{-+}_{d\sigma})(\omega) + 
({\bm G}^{-+}_{d\sigma} \circ {\bm G}^{a}_{d\sigma})(\omega)\right], \\
&{\bm \chi}^{-+}_{0nn\sigma}(\omega) =-i ({\bm G}^{-+}_{d\sigma} \circ {\bm G}^{+-}_{d\sigma})(\omega),
\end{align}
respectively.
We focus on the retarded and Keldysh components (see Appendix~\ref{app:GFs}) in the following discussions.
In the parameter regime under consideration, dominant contributions come from the zeroth mode of the correlation functions, which are denoted by
$\bar{\chi}^r_{0nn\sigma}(\omega) \equiv
\left( \chi^r_{0nn\sigma} \right)_{0}(\omega)$
and
$\bar{\chi}^K_{0nn\sigma}(\omega) \equiv
\left( \chi^K_{0nn\sigma} \right)_{0}(\omega)$.

In Fig.~\ref{fig:density_density}(a), we show the retarded component of the charge-charge correlation function, $\bar{\chi}^{r}_{0nn\sigma}(\omega)$, for $\epsilon_d=-U/2=-1$ and $V=0$.
The real (imaginary) part of $\bar{\chi}^{r}_{0nn\sigma}(\omega)$ is an even (odd) function with respect to $\omega$ (see Appendix~\ref{app:SymmetryRelation} for symmetry relations of various correlation functions), and has a dip (peak-and-dip) structure around $\omega = 0$.
These structures are reduced by the AC field due to the rise of the effective temperature discussed in Sec.~\ref{chap:noise_non}.

In Fig.~\ref{fig:density_density}(b), we show the Keldysh component of the charge-charge correlation function, $\bar{\chi}^{K}_{0nn\sigma}(\omega)$, for the same parameters.
The Keldysh component, $\bar{\chi}^{K}_{0nn\sigma}(\omega)$, is purely imaginary and even with respect to $\omega$.
In the absence of the AC field, the Keldysh component has two dip structures around $\omega = \pm \Delta/2$ due to the energy dependence of the spectral function.
The applied AC field increases the absolute value of the zero-frequency correlation function, $\bar{\chi}^K_{0nn}(0)$, reflecting the rise of the effective temperature.
The overall Lorentzian dip structure is insensitive to the AC field.

In the SCHF approximation, the dressed vertex functions include a RPA-type polarization function, which describes the dynamical screening effect.
The retarded and lesser components of the polarization function are obtained from Eq.~(\ref{eq:def of M}) as 
\begin{align}
\label{eq:retarded M}
&{\bm M}^r_{\sigma}(\omega)=\left[ {\bm 1}-U^2{\bm \chi}^r_{0nn \bar{\sigma}}(\omega){\bm \chi}^r_{0nn \sigma}(\omega) \right]^{-1}, \\
\label{eq:lesser M}
&{\bm M}^{-+}_{\sigma}(\omega) 
=U^2 {\bm M}^r_{\sigma} (\omega) \left[{\bm \chi}^r_{0nn \bar{\sigma}}(\omega){\bm \chi}^{-+}_{0nn \sigma}(\omega)\right. \nonumber \\
&\hspace{70pt} \left. +  {\bm \chi}^{-+}_{0nn \bar{\sigma}}(\omega){\bm \chi}^a_{0nn \sigma}(\omega) \right] {\bm M}^a_{\sigma}(\omega) ,
\end{align}
respectively.
We denote the diagonal elements of the retarded and Keldysh polarization functions by
$\bar{M}^r_{\sigma}(\omega) \equiv
\left( M^r_{\sigma} \right)_{0}(\omega)$
and
$\bar{M}^K_{\sigma}(\omega) \equiv
\left( M^K_{\sigma} \right)_{0}(\omega)$,
respectively.

Figure~\ref{fig:density_density}(c) and figure~\ref{fig:density_density}(d) show $\bar{M}^{r}_{\sigma}(\omega)$ and $\bar{M}^{K}_{\sigma}(\omega)$, respectively, for the same parameters.
The polarization functions have peak or dip structures around zero-frequency ($\omega = 0$), which come from those in the corresponding charge-charge correlation functions.
The peak of the real part of the retarded polarization function at $\omega=0$ is suppressed by the AC field, because the dynamical screening effect is weakened by the rise of the effective temperature.
The reduction of the Keldysh component of the polarization function occurs due to the same reason.
As the amplitude of the AC field increases, $\bar{M}^{r}_{\sigma}$ and $\bar{M}^{K}_{\sigma}$ approach 1 and 0, respectively.

\begin{figure*}[t]
\centering
\includegraphics*[width=0.8\hsize]{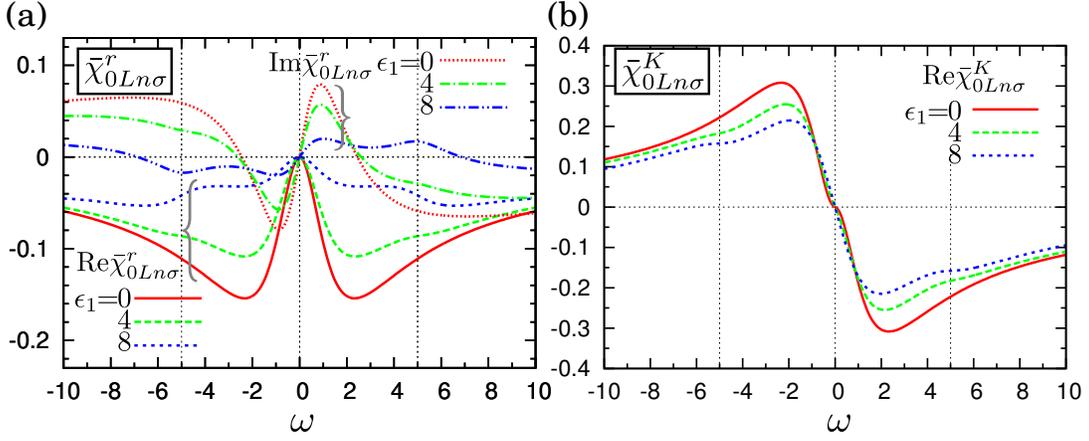}
\caption{\label{fig:current_charge}
(Color online)
(a) The retarded and (b) Keldysh components of the current-charge correlation function for $\epsilon_1=0$, $4$, and $8$.
Parameters are as follows: $\Delta_L=\Delta_R=1$, $\epsilon_d=-U/2=-1$, $\beta=20$, $V=0$, and $\Omega=5$.
}
\end{figure*}

We need to calculate the bare parts of the charge-current and current-charge correlation functions [see Eqs.~(\ref{eq:ncbare}) and (\ref{eq:cnbare})] to evaluate vertex corrections to the current noises.
Their retarded and lesser components are written in the Floquet representation as
\begin{widetext}
\begin{align}
&{\bm \chi}^{r}_{0\alpha n\sigma}(\omega)= 
i\Delta_{\alpha} 
[({\bm G}^{r}_{d\sigma} \circ {\bm G}^{r}_{d\sigma}{\bm f}_{\alpha})(\omega) - 
({\bm f}_{\alpha}{\bm G}^{a}_{d\sigma} \circ {\bm G}^{a}_{d\sigma})(\omega) + i 
{\bm \chi}^r_{0nn\sigma}(\omega) ],  \\
&{\bm \chi}^{-+}_{0\alpha n\sigma}(\omega) 
= -i\Delta_{\alpha} [{\bm G}^{-+}_{d\sigma} \circ {\bm G}^{r}_{d\sigma} 
\left( {\bm 1}-{\bm f}_{\alpha} \right) (\omega) +
\left( {\bm f}_{\alpha}{\bm G}^{a}_{d\sigma} \circ {\bm G}^{+-}_{d\sigma} \right) (\omega)]
-\Delta_{\alpha} {\bm \chi}^{-+}_{0nn\sigma}(\omega),\\
&{\bm \chi}^{r}_{0n\alpha\sigma}(\omega)= 
i\Delta_{\alpha} [ ({\bm G}^{r}_{d\sigma}{\bm f}_{\alpha} 
\circ {\bm G}^{a}_{d\sigma})(\omega) - ({\bm G}^{r}_{d\sigma} \circ {\bm f}_{\alpha}{\bm G}^{a}_{d\sigma})(\omega) ], \\
&{\bm \chi}^{-+}_{0n\alpha\sigma}(\omega)= - \left( {\bm \chi}^{-+}_{0\alpha n\sigma} \right)^{\dagger}(\omega).
\end{align}
\end{widetext}
The retarded and Keldysh components of the zeroth mode of the current-charge correlation function are denoted by
$\bar{\chi}^r_{0\alpha n\sigma}(\omega) \equiv
\left( \chi^r_{0\alpha n\sigma} \right)_{0}(\omega)$
and
$\bar{\chi}^{K}_{0\alpha n\sigma}(\omega) \equiv
\left( \chi^{K}_{0\alpha n\sigma} \right)_{0}(\omega)$,
respectively.

In Fig.~\ref{fig:current_charge}(a), we show the retarded component of the current-charge correlation function, $\bar{\chi}^{r}_{0L n\sigma}(\omega)$, for $\epsilon_d=-U/2=-1$, and $V=0$.
The real (imaginary) part of $\bar{\chi}^{r}_{0L n\sigma}(\omega)$ is an even (odd) function with respect to $\omega$.
The absolute value of $\bar{\chi}^{r}_{0L n\sigma}(\omega)$ tends to be reduced by the AC field because of the rise of the effective temperature, whereas the peaks develop at $\omega = \pm \Omega$ due to photon absorption and emission.
In Fig.~\ref{fig:current_charge}(b), we show the Keldysh component, $\bar{\chi}^{K}_{0L n\sigma}(\omega)$, for the same parameters.
The real part of $\bar{\chi}^{K}_{0L n\sigma}(\omega)$ is an odd function with respect to $\omega$.
The imaginary part vanishes at $V=0$.
The overall structure is insensitive to the AC field, while a small structure develops at $\pm \Omega$.

\subsection{Vertex corrections}

\begin{figure*}[t]
\centering
\includegraphics*[width=0.9\hsize]{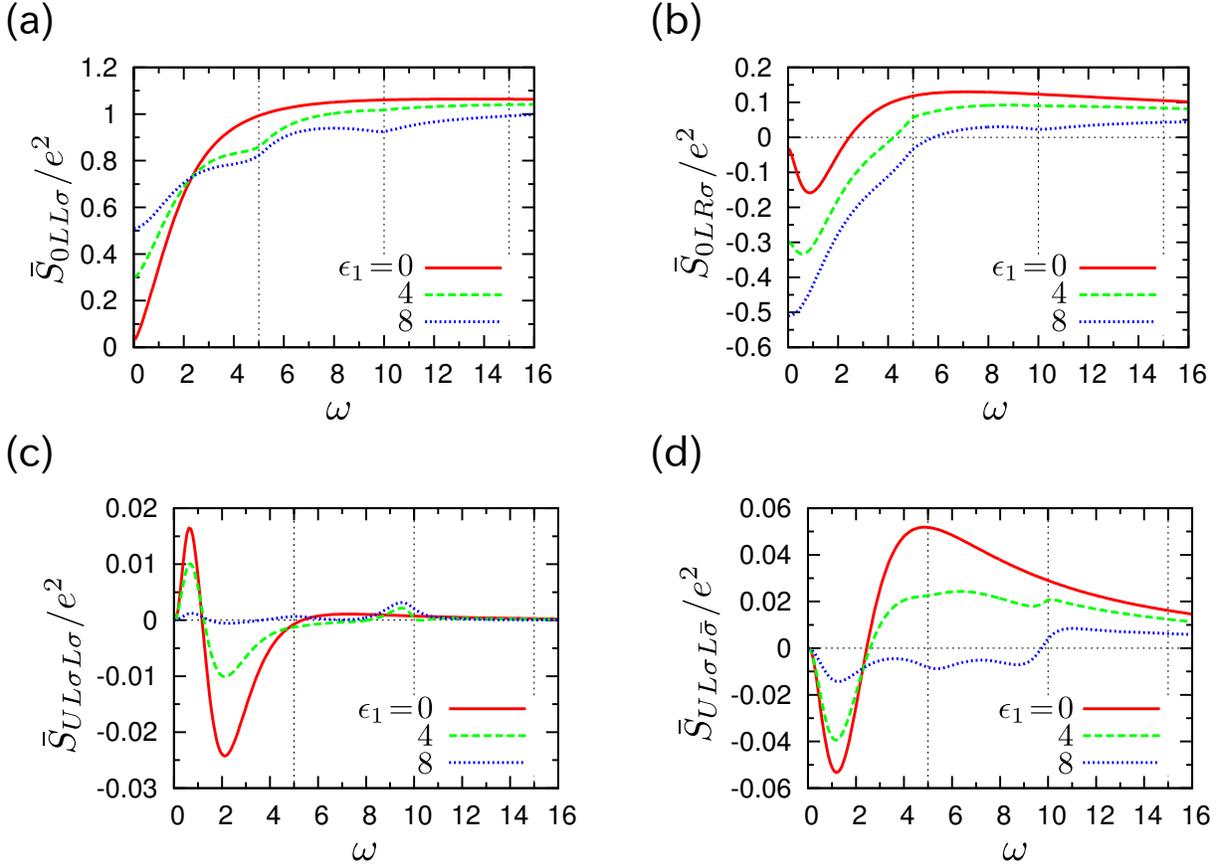}
\caption{\label{fig:Noise_spectrum_U}
(Color online)
The noise spectra for different amplitudes of the external field, $\epsilon_1=0$, $4$, and $8$.
The auto-correlation function, $\left(S_{0LL}\right)_{0}$, and the cross-correlation function, $\left(S_{0LR}\right)_{0}$, are shown in (a) and (b), respectively.
The vertex corrections to the current noise are displayed in the case of (c) parallel spins $\left(S_{0L\sigma L\sigma}\right)_{0}$ and (d) anti-parallel spins $\left(S_{0L\sigma L\bar{\sigma}}\right)_{0}$.
Parameters are as follows: $\Delta_L=\Delta_R=1$, $\epsilon_d=-U/2=-1$, $\beta=20$, $V=0$, and $\Omega=5$.
}
\end{figure*}

\begin{figure*}[t]
\centering
\includegraphics*[width=0.9\hsize]{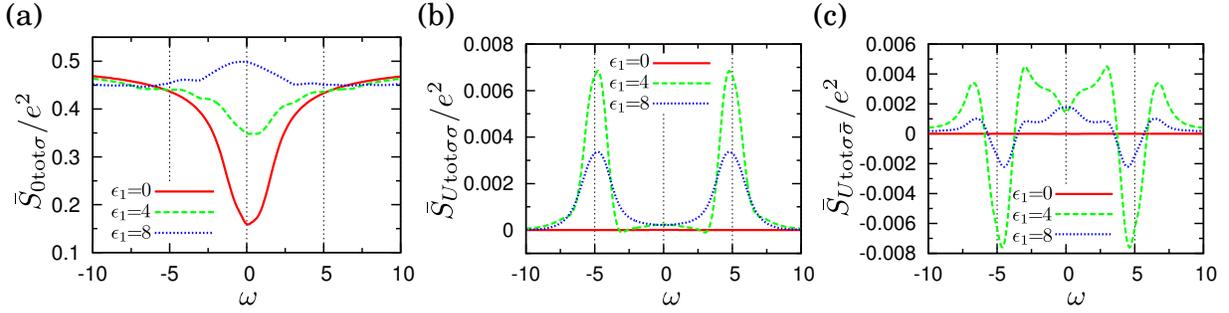}
\caption{\label{fig:Noise_spectrum_U_V_eps_total}
(Color online)
The noise spectra of the averaged current in the presence of a bias voltage ($V=2$) for different amplitudes of the external field, $\epsilon_1=0$, $4$, and $8$.
The bare part is shown in (a), while the vertex corrections for the parallel spins and the anti-parallel spins are shown in (b) and (c), respectively.
Parameters are as follows: $\Delta_L=\Delta_R=1$, $\epsilon_d=-U/2=-1$, $\beta=20$, $V=2$, and $\Omega=5$.
}
\end{figure*}

In this paper, we focus on the zeroth mode of the photon-assisted current noises, $\bar{S}_{0\alpha\alpha'\sigma}(\omega) \equiv (S_{0\alpha\alpha'\sigma})_{0}(\omega)$ and $\bar{S}_{U \alpha \sigma \alpha' \sigma'}(\omega) \equiv \left(S_{U \alpha \sigma \alpha' \sigma'} \right)_0 (\omega)$.
First, we consider the unbiased system ($V=0$), in which the relations
$\bar{S}_{0LL\sigma}(\omega)=\bar{S}_{0RR\sigma}(\omega)$ and 
$\bar{S}_{0LR\sigma}(\omega)=\bar{S}_{0RL\sigma}(\omega)$ hold.
The parallel-spin current noise is written as the sum of the bare term and the vertex correction term, $\bar{S}_{\alpha\sigma \alpha'\sigma}(\omega)=\bar{S}_{0\alpha\alpha'\sigma}(\omega)+\bar{S}_{U\alpha\sigma \alpha'\sigma}(\omega)$.
In contrast, the antiparallel-spin current noise is written only in terms of the vertex correction term, $\bar{S}_{\alpha\bar{\sigma} \alpha' \sigma}(\omega)=\bar{S}_{U\alpha\bar{\sigma} \alpha'\sigma}(\omega)$, because electrons with different spins cannot correlate with each other without the Coulomb interaction.
This indicates that we can directly evaluate the vertex correction, ${\bm S}_{U\alpha \bar{\sigma} \alpha' \sigma}(\omega)$, if the spin-dependent current noises can be measured using, for example, a spin filter.

The auto-correlation function of photon-assisted currents, $\bar{S}_{0LL\sigma}(\omega)$, is shown in Fig.~\ref{fig:Noise_spectrum_U}(a) for $\epsilon_d=-U/2=-1$ and $V=0$.
In the absence of the AC field, the auto-correlation function increases monotonically and reaches a finite value as the frequency increases.
This behavior is consistent with a previous study.~\cite{ding2013finite}
The AC field induces a step-like structure in the noise spectra, and each step starts to rise at integral multiples of the driving frequency, $\Omega=5$.
These features indicate that the present model behaves the same way as the multi-level QD system without the AC field~\cite{PhysRevB.84.235435} because of the appearance of the Floquet sidebands.~\cite{footnote3}
In contrast to the behavior of the zero-frequency noise, $\bar{S}_{0LL\sigma}(0)$, the auto-correlation function at high frequencies is suppressed by the AC field.
In Fig.~\ref{fig:Noise_spectrum_U}(b), we show the cross-correlation function, $\bar{S}_{0LR\sigma}(\omega)$, for the same parameters.
At zero frequency, the cross-correlation function is related to the auto-correlation function as $\bar{S}_{0LR\sigma}(0)=-\bar{S}_{0LL\sigma}(0)$, due to charge conservation.
In the absence of the AC field, $\bar{S}_{0LR\sigma}(\omega)$ is negative at low frequencies and positive at high frequencies.
The application of the AC field shifts $\bar{S}_{0LR\sigma}(\omega)$ in the negative direction and induces the dip structures at the integral multiples of the driving frequency.

In Fig.~\ref{fig:Noise_spectrum_U}(c) and \ref{fig:Noise_spectrum_U}(d), we show the vertex corrections, $S_{UL\sigma L\sigma'}(\omega)$, for the parallel spins ($\sigma'=\sigma$) and the anti-parallel spins ($\sigma'=\bar{\sigma}$), respectively. 
For $\epsilon_1=4$ and $8$, the peaks appear at integral multiples of the frequency of the AC field, and the spectra of the vertex corrections are strongly frequency-dependent due to the rich structure of the retarded current-charge correlation function [see Fig.~\ref{fig:current_charge}(a)].
On the other hand, the vertex corrections are suppressed on the whole by the application of the AC field, because the rise of the effective temperature weakens the dynamical screening effect.
These effects of the AC field on the vertex corrections are expected to be general in interacting electron systems.

To see effects of the bias voltage on the current noise, we consider the noise spectra associated with the averaged current, $(I_{\rm L}(t) - I_{\rm R}(t))/2$.
The bare and vertex correction parts of the averaged current noise are given by
${\bm S}_{0{\rm tot}\sigma}(\omega) \equiv ( {\bm S}_{0LL\sigma}(\omega) +{\bm S}_{0RR\sigma}(\omega) -{\bm S}_{0LR\sigma}(\omega) -{\bm S}_{0RL\sigma}(\omega) )/4$
and
${\bm S}_{U{\rm tot}\sigma\sigma'}(\omega) = ({\bm S}_{UL\sigma L\sigma'}(\omega)+{\bm S}_{UR\sigma R\sigma'}(\omega)-{\bm S}_{UL\sigma R\sigma'}(\omega)-{\bm S}_{UR\sigma L\sigma'}(\omega))/4$,
respectively.
In Fig.~\ref{fig:Noise_spectrum_U_V_eps_total}(a), we show the zeroth mode of the bare part of the noise associated with the averaged current, $\bar{S}_{0{\rm tot}\sigma}(\omega) \equiv (S_{0{\rm tot}\sigma})_{0}(\omega)$, for $\epsilon_d=-U/2=-1$ and $V=2$.
The three curves correspond to different amplitudes of the AC field; $\epsilon_1 = 0$, $4$, and $8$. 
The increase of the zero-frequency noise is due to the rise of the effective temperature.
In the high-frequency regime, the dependence of the noise on $\epsilon_1$ is weak, due to the cancellation of the auto- and cross-correlation functions.
Figure~\ref{fig:Noise_spectrum_U_V_eps_total}(b) and \ref{fig:Noise_spectrum_U_V_eps_total}(c) show the vertex correction parts,
$\bar{S}_{U{\rm tot}\sigma\sigma}(\omega) \equiv (S_{U{\rm tot}\sigma\sigma})_{0}(\omega)$ and
$\bar{S}_{U{\rm tot}\sigma\bar{\sigma}}(\omega) \equiv (S_{U{\rm tot}\sigma\bar{\sigma}})_{0}(\omega)$, respectively, for the same parameters.
Due to the assumed particle-hole symmetry, the vertex corrections associated with the correlation of the averaged current completely vanishes in the absence of the external AC field, which is consistent a the previous study.~\cite{ding2013finite}
However, the vertex corrections become finite when both the AC field and the bias voltage are simultaneously applied.
As seen in the figures, $\bar{S}_{U{\rm tot}\sigma\sigma}(\omega)$ 
($\bar{S}_{U{\rm tot}\sigma\bar{\sigma}}(\omega)$) has a clear peak (dip) at the driving frequency, $\Omega = 5$, which is caused by the currents resonantly driven by the external AC field.
This resonant character cannot be explained by the concept of the effective temperature.

\section{Conclusion}
\label{chap:conclusion}

In this paper, we investigated current noises through a QD system under an AC field to study effects that the Coulomb interaction had on photon-assisted current noises.
We employed the gauge-invariant formalism and the Floquet-GF method, and have derived explicit expressions for the vertex corrections of the photon-assisted current noises within the SCHF approximation.

We first focused on the zero-frequency photon-assisted current noise in Sec.~\ref{chap:noise_non}.
The bias-voltage dependence of the zero-frequency photon-assisted current noises has two features: singularities at the integral multiples of the frequency of the AC field and a remarkable enhancement of the zero-bias noise.
These two features are the direct consequences of the multi-step structure of the generalized distribution functions defined in Sec.~\ref{chap:PAT}.
An effective temperature was introduced as a good indicator of the enhancement of the zero-bias noise under the AC field.
The quantum coherence of electrons induced by the AC field manifests itself in the oscillatory behavior of the zero-frequency current noise as a function of the amplitude of the AC field.
These features are robust against the Coulomb interaction, and are in line with previous theoretical and experimental results on noninteracting electrons.

We next studied the frequency dependence of the photon-assisted current noises in Sec.~\ref{chap:noise_int}.
The AC field affects the vertex correction in two ways:
(1) The vertex correction is reduced by the AC field because the screening effect in the QD is suppressed by the rise of the effective temperature, and
(2) The frequency dependence of the vertex corrections shows resonant structures at the integer multiples of the driving frequency due to photon absorption and emission processes.
We expect that these are general features of interacting electron systems under a strong AC field.

Our study will offer a useful viewpoint for understanding photon-assisted transport of other phenomena, such as the Coulomb blockade~\cite{kaiser2007shot,wu2010noise,riwar2013zero} and the Kondo effect.~\cite{nordlander1999long,nordlander2000kondo,lopez2001low}
We finally point out that the present results serve as a starting point of the full counting statistics~\cite{levitov1996electron,ivanov1997coherent,RevModPhys.81.1665} of interacting electrons under time-dependent external fields.

\begin{acknowledgments}

We are grateful for helpful discussions with R. Sakano and T. Jonckheere.
This work was supported by JSPS KAKENHI Grant Numbers 24540316 and 26220711.
T.J.S. acknowledges financial support provided by the Advanced Leading Graduate Course for Photon Science (ALPS).

\end{acknowledgments}

\appendix

\section{Definitions of GFs on the real-time axis}
\label{app:GFs}

The GF of the QD electron is projected onto the real time axis by specifying the branches, and is denoted by $G^{\nu\nu'}_{d\sigma}(t,t')=G_{d\sigma}(\tau,\tau')$ for $\tau\in C^{\nu}$ and $\tau' \in C^{\nu'}$. 
The GFs $G^{-+}_{d\sigma}(t,t')$ and $G^{+-}_{d\sigma}(t,t')$ are called the lesser and greater GFs, respectively.
The retarded, advanced, and Keldysh component of the GF are defined as
\begin{eqnarray}
G^{r}_{d\sigma}(t,t') &\equiv& G^{--}_{d\sigma}(t,t') - G^{-+}_{d\sigma}(t,t'), \\
G^{a}_{d\sigma}(t,t') &\equiv& (G^{r}_{d\sigma}(t',t))^{*}, \\
G^{K}_{d\sigma}(t,t') &\equiv& G^{-+}_{d\sigma}(t,t')+G^{+-}_{d\sigma}(t,t'),
\end{eqnarray}
respectively.
All the GFs are determined by the retarded GF, $G^{r}_{d\sigma}(t,t')$, and the lesser GF, $G^{-+}_{d\sigma}(t,t')$.
The same relations hold for the other correlation functions.

\section{Symmetry Relations of the bare parts of correlation functions}
\label{app:SymmetryRelation}

From the explicit definitions of the correlation functions, we can prove the following symmetry relations:
\begin{align}
\left( \chi^{r}_{0nn\sigma} \right)_m(\omega) &= \left( \left( \chi^{r}_{0nn\sigma} \right)_{-m}(-\omega) \right)^{*}, \\
\left( \chi^{-+}_{0nn\sigma} \right)_m(\omega) 
&= \left( \chi^{+-}_{0nn\sigma} \right)_{m}(-\omega)  \nonumber \\
&= -\left( \left( \chi^{-+}_{0nn\sigma} \right)_{-m}(\omega) \right)^*,\\
\left( \chi^{r}_{0\alpha n\sigma} \right)_m(\omega) 
&= \left( \left( \chi^{r}_{0\alpha n\sigma} \right)_{-m}(-\omega) \right)^*,\\
\left( \chi^{-+}_{0\alpha n\sigma} \right)_m(\omega) 
&= - \left(\left( \chi^{+-}_{0\alpha n\sigma} \right)_{-m}(-\omega) \right)^{*}\nonumber \\
&= - \left( \left( \chi^{-+}_{0n\alpha \sigma} \right)_{-m}(\omega) \right)^*.
\end{align} 
From these relations, we can show that (1) the real (imaginary) part of $\bar{\chi}^{r}_{0nn\sigma}(\omega)$ is an even (odd) function of $\omega$, (2) $\bar{\chi}^{K}_{0nn\sigma}(\omega)$ is a purely imaginary even function of $\omega$, (3) the real (imaginary) part of $\bar{\chi}^{r}_{0\alpha n\sigma}(\omega)$ is an even (odd) function of $\omega$, and (4) the real (imaginary) part of $\bar{\chi}^{K}_{0\alpha n\sigma}(\omega)$ is an odd (even) function of $\omega$.

\bibliographystyle{unsrt}

\begin{thebibliography}{99}

\bibitem{PhysRev.129.647} P. K. Tien and J. P. Gordon, Phys. Rev. {\bf 129}, 647 (1963).
\bibitem{platero2004photon} G. Platero and R. Aguado, Phys. Rep. {\bf 395}, 1 (2004).
\bibitem{kohler2005driven} S. Kohler, J. Lehmann, and P. H{\"a}nggi, Phys. Rep. {\bf 406}, 379 (2005).


\bibitem{PhysRevLett.67.516} F. Grossmann, T. Dittrich, P. Jung, and P. H\"anggi, Phys. Rev. Lett. {\bf 67}, 516 (1991).

\bibitem{lesovik1994noise} G. B. Lesovik and L. S. Levitov, Phys. Rev. Lett. {\bf 72}, 538 (1994).


\bibitem{Pekola13} J. P. Pekola, O.-P. Saira, V. F. Maisi, A. Kemppinen, M. M\"ott\"onen, Y. A. Peshkin, and D. V. Averin,
Rev. Mod. Phys. {\bf 85}, 1421 (2013).

\bibitem{van2002electron}
W. G. van der Wiel, S. De Franceschi, J. M. Elzerman, T. Fujisawa, S. Tarucha, and L. P. Kouwenhoven,
Rev. Mod. Phys. {\bf 75}, 1 (2002).

\bibitem{feve2007demand} G. F{\`e}ve, A. Mah{\`e}, J.-M. Berroir, T. Kontos, B. Pla\c{c}ais, D. C. Glattli, A. Cavanna, B. Etienne, and Y. Jin, Science {\bf 316}, 1169 (2007).

\bibitem{dubois2013minimal} J. Dubois, T. Jullien, F. Portier, P. Roche, A. Cavanna, Y. Jin, W. Wegscheider, P. Roulleau, and D. C. Glattli, Nature {\bf 502}, 659 (2013).

\bibitem{ThierryReview} T. Martin, in {\it Nanophysics: Coherence and Transport},
edited by H. Bouchiat, Y. Gefen, S. Gu\'eron, G. Mon- tambaux, and J. Dalibard (Elsevier, 2004), Les Houches, Session LXXXI, p. 283;
arXiv:cond-mat/0501208.

\bibitem{blanter2000shot} Ya. M. Blanter and M. B{\"u}ttiker, Phys. Rep. {\bf 336}, 1 (2000).



\bibitem{hammer2011quantum} J. Hammer and W. Belzig, Phys. Rev. B {\bf 84}, 085419 (2011).
\bibitem{PhysRevLett.90.210602} S. Camalet, J. Lehmann, S. Kohler, and P. H\"anggi,
Phys. Rev. Lett. {\bf 90}, 210602 (2003).
\bibitem{chevallier2010detection} D. Chevallier, T. Jonckheere, E. Paladino, G. Falci, and T. Martin, Phys. Rev. B   {\bf 81}, 205411 (2010).





\bibitem{schoelkopf1998observation} R. J. Schoelkopf, A. A. Kozhevnikov, D. E Prober, and M. J. Rooks,
Phys. Rev. Lett. {\bf 80}, 2437 (1998).

\bibitem{kozhevnikov2000observation} A. A. Kozhevnikov, R. J. Schoelkopf, and D. E. Prober, Phys. Rev. Lett. {\bf 84}, 3398 (2000).

\bibitem{reydellet2003quantum} L.-H. Reydellet, P. Roche, D. C. Glattli, B. Etienne, and Y. Jin,
Phys. Rev. Lett. {\bf 90}, 176803 (2003).

\bibitem{gabelli2008dynamics} J. Gabelli and B. Reulet,
Phys. Rev. Lett. {\bf 100}, 026601 (2008).

\bibitem{gasse2013observation} G. Gasse, L. Spietz, C. Lupien, and B. Reulet,
Phys. Rev. B {\bf 88}, 241402 (2013).


\bibitem{Olkhovskaya08} S. Ol'khovskaya, J. Splettstoesser, M. Moskalets, and M. B\"uttiker, 
Phys. Rev. Lett. {\bf 101}, 166802 (2008).

\bibitem{bocquillon2013coherence} E. Bocquillon, V. Freulon, J.-M. Berroir, P. Degiovanni, 
B. Pla{\c{c}}ais, A. Cavanna, Y. Jin, G. F{\`e}ve, Science {\bf 339}, 1054 (2013).


\bibitem{Pedersen98} M. H. Pedersen and M. B\"uttiker, Phys. Rev. B {\bf 58}, 12993 (1998).


\bibitem{PhysRevB.46.7061} S. Hershfield, Phys. Rev. B {\bf 46}, 7061 (1992).


\bibitem{PhysRev.124.287} G. Baym and L. P. Kadanoff, Phys. Rev. {\bf 124}, 287 (1961).
\bibitem{PhysRev.127.1391} G. Baym, Phys. Rev. {\bf 127}, 1391 (1962).


\bibitem{ding2013finite} G. H. Ding and B. Dong, Phys. Rev. B {\bf 87}, 235303 (2013).


\bibitem{rech2012current} J. Rech, D. Chevallier, T. Jonckheere, and T. Martin,
Phys. Rev. B {\bf 85}, 035419 (2012).

\bibitem{tsuji2008correlated} N. Tsuji, T. Oka, and H. Aoki, Phys. Rev. B {\bf 78}, 235124 (2008).




\bibitem{footnote1} The vector potentials are introduced by the Peierls substitution as $t_{\alpha}e^{ie\int_{\alpha} dx A_{\sigma}(x)}$ where $\int_{\alpha} dx A_\sigma(x)$ is the integral of the vector potentials from lead $\alpha$ to the QD.
The positive currents flow from lead $\alpha$ to the QD.
In this paper, the integral is averaged by using the length between the lead and the QD, which is set one.

\bibitem{chou1985equilibrium} K. Chou, Z. Su, B. Hao, and L. Yu,
Phys. Rep. {\bf 118}, 1 (1985).

\bibitem{kamenev2011field} A. Kamenev, {\it Field theory of non-equilibrium systems} (Cambridge University Press, Cambridge, 2011).


\bibitem{PhysRev.118.1417} J. M. Luttinger and J. C. Ward, Phys. Rev. {\bf 118}, 1417 (1960).

\bibitem{PhysRevD.10.2428} J. M. Cornwall, R. Jackiw, and E. Tomboulis,
Phys. Rev. D {\bf 10}, 2428 (1974).

\bibitem{footnote2} In the fermionic system, the expectation value of the fermion field vanishes.

\bibitem{footnote3}
The present formulation is equivalent, up to the gauge transformation, to the case where the effect of the AC field is incorporated only in the dot GFs.
In the latter picture, the spectral function of the dot GF has peaks at the integral multiples of the external field frequency, which are called the Floquet sidebands.


\bibitem{haug2008quantum} H. Haug, A.-P. Jauho,
{\it Quantum kinetics in transport and optics of semiconductors}
(Springer, New York, 2008).

\bibitem{PhysRevB.73.075108} V. M. Turkowski and J. K. Freericks, Phys. Rev. B {\bf 73}, 075108 (2006).



\bibitem{caso2012defining} A. Caso, L. Arrachea, and G. S. Lozano,
Eur. Phys. J. B {\bf 85}, 266 (2012).
\bibitem{casas2003temperature} J. Casas-V{\'a}zquez and D. Jou,
Rep. Prog. Phys. {\bf 66}, 1937 (2003).



\bibitem{PhysRevB.84.235435} N. Gabdank, E. A. Rothstein, O. Entin-Wohlman, and A. Aharony,
Phys. Rev. B {\bf 84}, 235435 (2011).



\bibitem{kaiser2007shot} F. J. Kaiser and S. Kohler, Ann. Phys. {\bf 16}, 702 (2007).

\bibitem{wu2010noise} B. H. Wu and C. Timm, Phys. Rev. B {\bf 81}, 075309 (2010).

\bibitem{riwar2013zero} R.-P. Riwar, J. Splettstoesser, and J. K\"onig, Phys. Rev. B {\bf 87}, 195407 (2013).


\bibitem{nordlander1999long} P. Nordlander, M. Pustilnik, Y. Meir, N. S. Wingreen, and D. C. Langreth, Phys. Rev. Lett {\bf 83}, 808 (1999).

\bibitem{nordlander2000kondo} P. Nordlander, N. S. Wingreen, Y. Meir, and D. C. Langreth, Phys. Rev. B {\bf 61}, 2146 (2000).

\bibitem{lopez2001low} R. L{\'o}pez, R. Aguado, G. Platero, and C. Tejedor, Phys. Rev. B {\bf 64}, 075319 (2001).

\bibitem{levitov1996electron}
L. S. Levitov, H. W. Lee, and G. B. Lesovik, J. Math. Phys. {\bf 37}, 4845 (1996).

\bibitem{ivanov1997coherent} D. A. Ivanov, H. W. Lee, and L. S. Levitov, Phys. Rev. B {\bf 56}, 6839 (1997).

\bibitem{RevModPhys.81.1665} M. Esposito, U. Harbola, and S. Mukamel,
Rev. Mod. Phys. {\bf 81}, 1665 (2009).





\end{thebibliography}

\end{document}